# A Market-Clearing-based Sensitivity Model for Locational Marginal and Average Carbon Emission

Zelong Lu, *Student Member, IEEE*, Lei Yan, Jianxue Wang, *Senior Member, IEEE*, Chongqing Kang, *Fellow, IEEE*, Mohammad Shahidehpour, *Life Fellow, IEEE*, Zuyi Li, *Senior Member, IEEE*

*Abstract--* This letter proposes a market-clearing-based locational marginal carbon emission (LMCE) metric to assess the marginal carbon emission effect of nodal load demand. Unlike the prevalent carbon emission flow (CEF) method that relies on a hypothetical power-flow tracking process, the proposed LMCE metric depends on a novel sensitivity analysis of market-clearing results, capable of revealing both energy-dependent and network-dependent impacts on emissions. Additionally, we introduce a locational average carbon emission (LACE) metric, derived from LMCE, to effectively measure the general emission effect. It offers insights into demand-side carbon emission effects, such as a negative LMCE and LACE indicating emission reduction even as load increases. It can also prevent excessive demand-side emission allocations. Overall, the proposed method provides a clear perspective for the ongoing decarbonization policies.

## I. INTRODUCTION

THE electricity consumption of demand-side plays a vital role in reducing carbon emissions from electricity producers [1]. Recent studies on assessing demand-side carbon emission can be divided into two types: 1) Average emission factors, which measure user-side carbon emission obligation via the wide-spread carbon emission flow (CEF) method [1]; 2) Marginal emission factors, revealing overall emissions' variations with slight changes in electricity demand [2], [3].

Noticed that the CEF model relies on a hypothetical power-flow tracking process under a proportional sharing principle, which could face practical challenges in market environments. A basic example is provided in Appendix A presented in [4]. Meanwhile, the concept of nodal carbon emission intensity (NCI) is usually non-negative [1]. Yet, finding a way to calculate a negative NCI, showing how increasing flexible loads can lower carbon emissions even, remains unresolved.

Currently, marginal factors are based on sensitivity analysis within a linear programming (LP) framework. However, this approach may suffer from numerical issues in diverse market conditions, since the objective function is set as linear and only focuses on binding constraints [2], [5]. Moreover, marginal emission factors provide information for small changes, not for large changes [5]. The direct use of marginal emission factors could involve excessive allocation. Additionally, it is still an open question on how to design a transparent incentive to quantify the impact of market-clearing results and network operating conditions on the marginal emission factors. Hence, there is a growing need for a novel emission measuring metric to quantify and validate the demand-side emission reductions marginally and on average by peak load reduction and renewable energy generation in market-clearing frameworks.

In this letter, we propose a market-clearing-based locational marginal carbon emission (LMCE) metric to provide a clear and fair incentive for the demand-side decarbonization, which can be decomposed into energy-dependent and network-dependent components. Furthermore, a locational average carbon emission (LACE) metric is derived based on the LMCE, which can reveal the negative NCI results and eliminate excessive allocation of carbon emissions to the demand side.

## II. LOCATIONAL CARBON EMISSION METRICS

### A. Market-clearing-based sensitivity analysis to derive LMCE

In this section, we propose the LMCE metric via a market-clearing-based sensitivity analysis based on perturbation technique, similar to the one used in [5]. Our method is also applicable to cases where the objective function is quadratic or the flow constraints are nonlinear (see Appendix B in [4]). The basic idea is described as follows. For simplicity, the market-clearing problem is still abstracted as a LP form:

$$\min \quad c^T x \quad (1a)$$
$$\text{s.t.} \quad Ax = b: \pi \quad (1b)$$
$$v \leq Bx \leq u: \phi, \psi \quad (1c)$$

The objective function is to minimize the bid-based cost for electricity market. The decision variables $x$ include the generation dispatch. The constraints $Ax = b$ and $v \leq Bx \leq u$ represent various operating constraints, such as power balance, line flow limits, and individual unit constraints. The line flow limits as part of inequality constraints can be further represented by a classical power transfer distribution factor (PTDF) model:

$$-\bar{P}_L \leq PTDF \times [P_g - P_d] \leq \bar{P}_L \quad (2)$$

where the decision variable vector $x$ includes $P_g$ and $P_d$ is the constant vector of the load demand included in $b$. It is evident that the lower and upper bounds of (1c) are functions of load demand; so (1c) can be reformulated as

$$v(b) \leq Bx \leq u(b): \phi, \psi \quad (3)$$

Note that $\pi$ in (1b), $\psi$ and $\phi$ in (3) are vectors of Lagrangian multipliers, respectively.

By differentiating the corresponding Karush-Kuhn-Tucker (KKT) conditions for (1) and (3) with respect to $b$, we get:

$$\begin{bmatrix} 0 & A & 0 & 0 \\ A^T & 0 & -B^T & B^T \\ 0 & \Psi B & -W_1 & 0 \\ 0 & \Phi B & 0 & W_2 \end{bmatrix} \cdot \begin{bmatrix} \partial \pi / \partial b \\ \partial x / \partial b \\ \partial \psi / \partial b \\ \partial \phi / \partial b \end{bmatrix} = \begin{bmatrix} 1 \\ 0 \\ \Psi * \frac{\partial u(b)}{\partial b} \\ \Phi * \frac{\partial v(b)}{\partial b} \end{bmatrix} \quad (4)$$

where $w_1$ and $w_2$ are the slack vectors of (3); $W_1 = \text{Diag}(w_1)$, $W_2 = \text{Diag}(w_2)$, $\Psi = \text{Diag}(\psi)$, and $\Phi = \text{Diag}(\phi)$. $\mathbf{1}$ is a unity matrix and $\mathbf{0}$ is a zero matrix.

Denote the coefficient matrix on the left-hand side of (4) as $H$, the column vector on the right-hand side of (4) as $d\rho$, and $dz = [\partial \pi / \partial b, \partial x / \partial b, \partial \psi / \partial b, \partial \phi / \partial b]^T$ Then, $H \cdot dz = d\rho$. $H$ in (4) is highly sparse where sparsity techniques can be employed to calculate $\partial x / \partial b$, which is of interest for calculating LMCE. Considering potential numerical issues in market-clearing-based sensitivity methods, matrix $H$ might have a lower rank than its variable vector or be in a degenerate state with binding constraints but null multipliers [5]. In these cases, $H$ could be an ill-conditioned matrix. Consequently, the singular value decomposition (SVD) technique is applied.

$$d\mathbf{z} = \mathbf{V} \cdot \mathbf{\Sigma}^+ \cdot \mathbf{U}^T \cdot d\boldsymbol{\rho} \quad (5)$$

where $\mathbf{H} = \mathbf{U} \cdot \mathbf{\Sigma} \cdot \mathbf{V}^T$. $\mathbf{U}$ and $\mathbf{V}^T$ are orthogonal matrices spanning the columns and rows of $\mathbf{H}$, respectively. $\mathbf{\Sigma}$ is a diagonal matrix with the singular values of $\mathbf{H}$ on its diagonal and $\mathbf{\Sigma}^+$ is the pseudoinverse of $\mathbf{\Sigma}$. Specifically, for each non-zero element $\sigma_i$ of $\mathbf{\Sigma}$, its value in $\mathbf{\Sigma}^+$ is $1/\sigma_i$, and for zero elements in $\mathbf{\Sigma}$, they remain zero in $\mathbf{\Sigma}^+$.

The total carbon emission across the system depends on the decision variable $\mathbf{x}$. That is,

$$f(\mathbf{x}) \triangleq \sum_{g \in S_G} \sum_t k_g * p_{g,t}^* = \mathbf{K}^T \cdot \mathbf{x} \quad (6)$$

Given the implicit function (4), LMCE is defined as:

$$\boldsymbol{\ell}^{LMCE} = \frac{\partial f(\mathbf{x})}{\partial \mathbf{x}} \frac{\partial \mathbf{x}}{\partial \mathbf{b}} = \mathbf{K}^T \cdot [\mathbf{0}, \mathbf{I}, \mathbf{0}, \mathbf{0}] \cdot \mathbf{V} \cdot \mathbf{\Sigma}^+ \cdot \mathbf{U}^T \cdot d\boldsymbol{\rho} \quad (7)$$

LMCE represents the incremental change of total system emission in response to a unit load change at a specific location. Since LMCE definition is similar to that of LMP, **Lemma 1** provides a clear and intuitive analysis of the physical attributes of LMCE, which is inspired by the LMP decomposition, characterizing the factors affecting nodal carbon emissions.

**Lemma 1.** *LMCE can be decomposed into energy-dependent and network-dependent components which are formulated as:*

$$\boldsymbol{\ell}^{LMCE} = \frac{\partial \text{TE}}{\partial \mathbf{b}} = \frac{\partial f(\mathbf{x})}{\partial \mathbf{b}} = \frac{\partial f(\mathbf{x})}{\partial \mathbf{x}} \frac{\partial \mathbf{x}}{\partial \mathbf{b}}$$

$$= \mathbf{K}^T \cdot \underbrace{[\mathbf{0}, \mathbf{I}, \mathbf{0}, \mathbf{0}] \cdot \mathbf{V} \cdot \mathbf{\Sigma}^+ \cdot \mathbf{U}^T}_{Y} d\boldsymbol{\rho} = \mathbf{K}^T \cdot \mathbf{Y} \cdot \begin{bmatrix} 1 \\ 0 \\ \mathbf{\Psi} * \frac{\partial u(b)}{\partial b} \\ \mathbf{\Phi} * \frac{\partial v(b)}{\partial b} \end{bmatrix}$$

$$= \mathbf{K}^T \cdot \mathbf{Y} \cdot \begin{bmatrix} 1 \\ 0 \\ 0 \\ 0 \end{bmatrix} + \mathbf{K}^T \cdot \mathbf{Y} \cdot \begin{bmatrix} 0 \\ 0 \\ \mathbf{\Psi} * \frac{\partial u(b)}{\partial b} \\ \mathbf{\Phi} * \frac{\partial v(b)}{\partial b} \end{bmatrix} = \boldsymbol{\ell}_{energy}^{LMCE} + \boldsymbol{\ell}_{network}^{LMCE}$$

Lemma 1 quantifies the impacts of market-clearing results and network operating conditions on the LMCE metric. Hence, the LMCE method offers a clearer incentive for market participants as compared with the CEF model.

### B. Riemann-sum based LACE metric

Similar to conventional LMP metrics, using LMCE for carbon emission measurement may lead to overestimating demand-side carbon emissions. Hence, we propose a LACE metric to ensure that the demand-side share of carbon emission is equal to the actual production of carbon emission on the generation side.

Using LMCE, the marginal emission can be denoted as the system-wide carbon emission change driven by a slight change in nodal load demand. Assume that $b_{i,t}$ is the total energy consumption of node $i$ at time $t$ and $\mathbf{b}$ represents the vector of $b_{i,t}$. Since the market-clearing results $\mathbf{x}$ can be taken as a function of load demand $\mathbf{b}$, LMCE can also be regarded as a function of $\mathbf{b}$. Hence,

$$\boldsymbol{\ell}^{LMCE}(\mathbf{b}) = \frac{\partial f(\mathbf{x})}{\partial \mathbf{b}} = \frac{\partial f(\mathbf{x})}{\partial \mathbf{x}} \frac{\partial \mathbf{x}(\mathbf{b})}{\partial \mathbf{b}} = \mathbf{K}^T \cdot \frac{\partial \mathbf{x}(\mathbf{b})}{\partial \mathbf{b}} \quad (8)$$

LACE can be calculated as a Riemann-sum of LMCE, which is formed as the uniform average of the marginal carbon emission along the segment of nodal load demand from $\mathbf{0}$ to $\mathbf{b}^*$. For a given nodal load $b_{i,t}^*$, LACE is shown as:

$$\ell_{i,t}^{LACE} \triangleq \int_0^1 \ell_{i,t}^{LMCE}(\sigma \mathbf{b}^*) d\sigma / b_{i,t}^*$$
$$= \int_0^1 \mathbf{K}^T \cdot \frac{\partial \mathbf{x}(\sigma \mathbf{b}^*)}{\partial b_{i,t}} d\sigma / b_{i,t}^* \quad (9)$$

**Lemma 2.** *For any polynomial $f(\mathbf{x})$, the LACE metric ensures that the total carbon emission created on the generation side is the same as that allocated to the demand side*, i.e., $\sum_t \sum_i \ell_{i,t}^{LACE} * b_{i,t}^* = f(\mathbf{x}(\mathbf{b}^*))$.

**Proof**: Since the original problem (1) is convex and feasible for $\mathbf{b} = \mathbf{0}$ and $\mathbf{b} = \mathbf{b}^*$, it is also feasible for $\mathbf{b} = \sigma \mathbf{b}^*, \forall \sigma \in [0,1]$. The total demand-side carbon emission can be regarded as a function of load demand $\mathbf{b}$, i.e, $f(\mathbf{b})$. Further, we can obtain the following result by the Newton-Leibniz theorem:

$$f(\mathbf{b}^*) - f(\mathbf{b}^* - \triangle \mathbf{b}^*) = \int_{\mathbf{b}^* - \triangle \mathbf{b}^*}^{\mathbf{b}^*} \frac{\partial f(\mathbf{b})}{\partial \mathbf{b}} d\mathbf{b}$$
$$= \sum_i \sum_t \int_{1-\varepsilon}^{1} \ell_{i,t}^{LMCE}(\sigma \mathbf{b}^*) d\sigma \quad (10)$$

Then, we have:

$$f(\mathbf{b}^*) - f(\mathbf{0}) = f(\mathbf{b}^*) = \sum_i \sum_t \int_0^1 \ell_{i,t}^{LMCE}(\sigma \mathbf{b}^*) d\sigma$$
$$= \sum_i \sum_t \ell_{i,t}^{LACE} * b_{i,t}^* \quad (11)$$

Lemma 2 ensures the fairness of LACE in allocating the demand-side emission without any excess.

**Lemma 3:** *When the polynomial $f(\mathbf{x})$ is linear, as the segment of nodal load changes from 0 to $b_{i,t}^*$, the corresponding LMCE metric $\ell_{i,t}^{LMCE}$ will remain constant within a finite number of intervals. Then, the nodal LACE can be formulated as:*

$$\ell_{i,t}^{LACE} = \left[ \sum_{m=1}^{M} (y_m - y_{m-1}) * \ell_{i,t,m-1}^{LMCE} \right] \quad (12)$$

**Proof:** According to the multiparametric linear programming theory, the marginal value follows a finite number *M* of critical region, which are thoroughly investigated in [6].

Lemma 3 offers an acceleration method for the calculation of LACE when the stepwise LMCE curve is available.

### III. CASE STUDY

To illustrate the LMCE and LACE model and highlight their differences from the CEF model, a simple 3-bus example is offered in this section. Fig. 1 shows the carbon emission distribution calculated based on the CEF model, including the NCI and the branch carbon emission intensity (BCI) following an equal proportional sharing principle [1]. Parameters are detailed in Appendix C as presented in [4]. According to the flow-tracing principle of the CEF model, the NCIs at Buses 1-3 are 0.2, 0.2, and 0.32 tCO2/MWh, respectively.

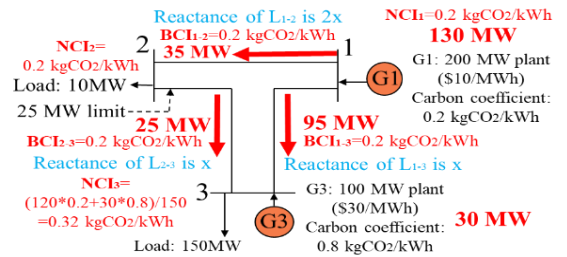

Fig. 1 Carbon emission distribution by the CEF model

In comparison, the LMCEs at those buses are 0.2, -1, and 0.8 tCO2/MWh, respectively. The results can be verified as follows.

Based on the above 3-bus example, we set the results in Fig. 1 as the base case. Table I provides the carbon emission results calculated by the proposed market-clearing-based sensitivity method. Based on the concept of LMCE, a unit load increase at different nodes changes the optimal market clearing, which in turn affects the system's total carbon emission.

TABLE I. Change in the system-wide carbon emissions caused by an increase of unit load demand at different nodes

|  | Gen1 (MW) | Gen3 (MW) | Total $CO_2$ (t$CO_2$) | LMCE (t$CO_2$/MWh) |
|---|---|---|---|---|
| Base case | 130 | 30 | 50.0 | \ |
| Add 1 MW to Bus 1 | 131 | 30 | 50.2 | 0.2 |
| Add 1 MW to Bus 2 | 133 | 28 | 49.0 | -1.0 |
| Add 1 MW to Bus 3 | 130 | 31 | 50.8 | 0.8 |

It is interesting to note that the proposed method can reveal a negative LMCE without a non-negative limitation of the CEF model. This also verifies that the proposed method can measure various nodal carbon reduction effects. However, as stated in Section II.C, directly using the LMCE metric may result in excessive allocation of carbon emission to demand-side. As shown in Table I, the total carbon emission allocated to the demand-side is 110 t$CO_2$ (-1 t$CO_2$/MWh*10 MWh+0.8 t$CO_2$/MWh*150 MWh), while the actual system-wide carbon emission is 50 t$CO_2$. Hence, we further verify the use of the proposed LACE metric. Based on Lemma 3, the nodal carbon intensity for Bus 2 is -0.07692 kg$CO_2$/kWh, as shown in Fig. 2.

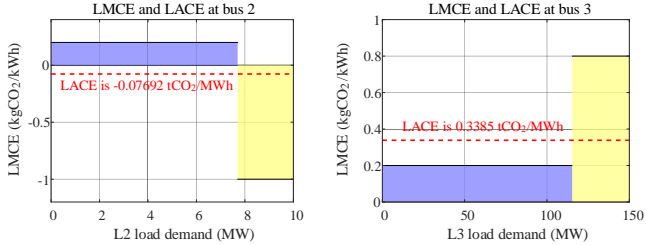

Fig. 2 LACE calculation by the proposed method

Table II compares the carbon emission allocations of CEF and LCME models. Here, the load at Bus 2 gets 0.77 t$CO_2$ credit due to its contribution to the marginal reduction of carbon emission. Note that although the value of the LACE differs from the NCI calculated by the CEF model, the allocated demand-side carbon emission aligns perfectly with the actual system-side carbon emission (-0.07692 t$CO_2$/MWh*10 MWh+0.3385 t$CO_2$/MWh*150 MWh=50 t$CO_2$). It verifies the validity of LACE, which can effectively allocate carbon emission to demand side and reveal the average nodal carbon reduction effect. It also means that there are different ways of achieving balanced carbon emission allocation. While the CEF model relies on the assumed equal proportional sharing principle, the LMCE model is based on a rigorous process that only relies on market clearing results.

TABLE II. Comparison of carbon allocation model

| Bus | Load (MW) | CEF Model | | LCME Model | |
|---|---|---|---|---|---|
| | | NCI (t$CO_2$/MWh) | Carbon Allocation (t$CO_2$) | LACE (t$CO_2$/MWh) | Carbon Allocation (t$CO_2$) |
| 2 | 10 | 0.20 | 2 | -0.07692 | -0.77 |
| 3 | 150 | 0.32 | 48 | 0.3385 | 50.77 |
| Total | 160 | \ | 50 | \ | 50 |

Furthermore, to demonstrate the relationship between the cleared LMCE and the network operation condition, we modify an IEEE 6-bus system over 24 hours to verify the solution offered by the proposed LMCE method. The spatial-temporal distribution of the power flow is provided in Fig. 3(a) and the clearing result for the LMCE is shown in Fig. 3(b). When there is no congestion, i.e., the line flow rate (absolute value of line flow over the line's rated capacity) is less than 100%, the nodal LMCEs are all the same which are determined by the emission coefficient of the marginal unit.

When there is a line congestion (represented by black blocks in Fig. 3(a)), LMCE will vary with nodes. As shown in Fig. 3(b), in the congested hours circled in red, the LMCE is affected by the congestion in the system, while the change of load at different nodes will have varying effect on the carbon emission of the entire system.

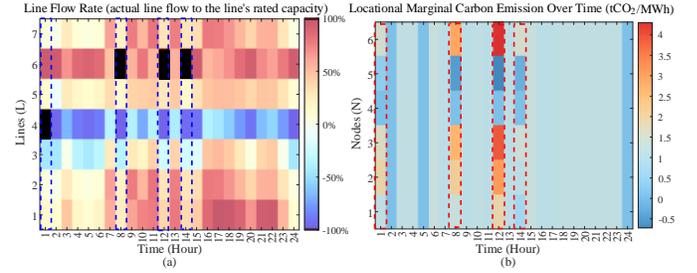

Fig. 3 Spatial-temporal distribution of line flow and LMCE

To further illustrate Lemma 1, Appendix D presented in [4] provides the decomposition results of LMCE as energy-dependent and network-dependent components. The results indicate that the proposed LMCE solution depends on market-clearing outcomes and network operation conditions. Thus, it provides a clearer incentive for market participants to adjust their loads compared with the conventional CEF model.

## IV. CONCLUSION

This letter proposes a decomposable LMCE metric via a market-clearing-based sensitivity analysis method. Moreover, a novel LACE model is developed to eliminate excessive carbon emission allocation in the demand side. In our next steps, we will expand the LMCE and LACE model to integrated energy systems, while also exploring how to fully motivate demand responses to participate in China's certified emission reduction market via the extension of the proposed methods.

## REFERENCES

[1] C. Kang, T. Zhou, Q. Chen, J. Wang, Y. Sun, Q. Xia, and H. Yan, "Carbon emission flow from generation to demand: A network-based model," *IEEE Transactions on Smart Grid*, vol. 6, no. 5, pp. 2386–2394, 2015.
[2] L. F. Valenzuela, A. Degleris, A. E. Gamal, M. Pavone, and R. Rajagopal, "Dynamic locational marginal emissions via implicit differentiation," *IEEE Transactions on Power Systems*. early access, 2023.
[3] B. Park, J. Dong, B. Liu, and T. Kuruganti, "Decarbonizing the grid: Utilizing demand-side flexibility for carbon emission reduction through locational marginal emissions in distribution networks," *Applied Energy*, vol. 330, p. 120303, 2023.
[4] Z. Lu. "Appendix for A Market-clearing-based Sensitivity Model for Locational Marginal Carbon Emission," https://bit.ly/Appendix_LMCE.
[5] A. J. Conejo, E. Castillo, R. Minguez and F. Milano, "Locational marginal price sensitivities," *IEEE Transactions on Power Systems*, vol. 20, no. 4, pp. 2026-2033, Nov. 2005.
[6] G. B. Dantzig and M. N. Thapa, Linear Programming 2: Theory and Extensions. Springer, 2003.

# A Market-Clearing-based Sensitivity Model for Locational Marginal and Average Carbon Emission

Zelong Lu, *Student Member, IEEE*, Lei Yan, Jianxue Wang, *Senior Member, IEEE*, Chongqing Kang, *Fellow, IEEE*, Mohammad Shahidehpour, *Life Fellow, IEEE*, Zuyi Li, *Senior Member, IEEE*

## APPENDIX A

Given that the CEF model relies on an equal proportional sharing principle, nodal carbon intensities would differ for various load aggregator (LA) organization forms under the same market-clearing outcome.

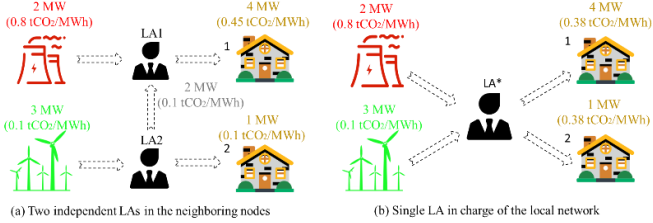

Fig.A.1 A base case to illustrate the unfair incentive of the CEF method

In Fig. A.1 (a), two LAs are in charge of distributed loads within their service areas. Under the CEF method, the nodal carbon emission intensity for LA1 is 0.45 tCO$_2$/MWh with a 4MW load while that for LA2 is 0.1 tCO$_2$/MWh with a 1MW load. However, if we the two LAs are combined into a single LA in charge of all loads, all nodes share the same carbon emission intensity of 0.38 tCO$_2$/MWh. This makes the CEF model's application in market settlement debatable, potentially hindering the motivation for the demand-side decarbonization.

## APPENDIX B

For simplicity, the mathematical problem can be extracted as a SOCP form in the distribution-level network:

$$\min \quad c^T x$$
$$\text{s.t.} \quad Ax = b : \pi$$
$$L(b) \leq Fx \leq U(b) : \phi, \psi$$
$$\|G_i x + h_i\|_2 \leq D_i x + e_i : \|\tau_i\|_2 + \beta_i = \kappa_i \text{ for all } i \quad \text{(B1)}$$

The differentiation of the KKT conditions can be formed as:

$$\begin{bmatrix} 0 & A & 0 & 0 & 0 & 0 & 0 \\ A^T & 0 & D^T & G^T & -F^T & F^T & 0 \\ 0 & \Psi * F & 0 & 0 & -W_1 & 0 & 0 \\ 0 & \Phi * F & 0 & 0 & 0 & W_2 & 0 \\ 0 & K * D & 0 & DE & 0 & 0 & 0 \\ 0 & B*G/\|Gx+h\|_2 - D & 0 & 0 & 0 & 0 & -W_3 \\ 0 & 0 & 1/\|\tau\|_2 & 0 & -I & 0 & I \end{bmatrix}$$

$$* \begin{bmatrix} \partial\pi/\partial b \\ \partial x/\partial b \\ \partial\tau/\partial b \\ \partial\kappa/\partial b \\ \partial\psi/\partial b \\ \partial\phi/\partial b \\ \partial\beta/\partial b \end{bmatrix} = \begin{bmatrix} 1 \\ 0 \\ \Psi * \frac{\partial U(b)}{\partial b} \\ \Phi * \frac{\partial L(b)}{\partial b} \\ 0 \\ 0 \\ 0 \end{bmatrix} \quad \text{(B2)}$$

Similarly, the quadratic objective function problem with linear constraints could be abstracted as below.

$$\min \quad \frac{1}{2} x^T Q x + c^T x + b$$
$$\text{s.t.} \quad Ax = b : \pi$$
$$v(b) \leq Bx \leq u(b) : \phi, \psi \quad \text{(B3)}$$

By differentiating the corresponding KKT conditions with respect to $b$, we get:

$$\begin{bmatrix} 0 & A & 0 & 0 \\ A^T & -Q & -B^T & B^T \\ 0 & \Psi B & -W_1 & 0 \\ 0 & \Phi B & 0 & W_2 \end{bmatrix} \cdot \begin{bmatrix} \partial\pi/\partial b \\ \partial x/\partial b \\ \partial\psi/\partial b \\ \partial\phi/\partial b \end{bmatrix} = \begin{bmatrix} 1 \\ 0 \\ \Psi * \frac{\partial u(b)}{\partial b} \\ \Phi * \frac{\partial v(b)}{\partial b} \end{bmatrix} \quad \text{(B4)}$$

## APPENDIX C

Fig. 2 displays the carbon emission distribution calculated based on the CEF model. Generator 1 (G1) has a capacity of 200MW, with a bid price of \$10/MWh and a carbon emission factor of 0.2 kgCO$_2$/kWh. Generator 3 (G3) has a capacity of 100MW, with a bid price of \$30/MWh and a carbon emission factor of 0.8 kgCO$_2$/kWh. The reactance of Line 1-2 is twice that of Line 2-3. The reactance of Line 2-3 and that of Line 1-3 are the same. The capacity limit of Line 2-3 is 25 MW. According to the flow-tracing principle of the CEF model, the NCIs of Bus 1, Bus 2, and Bus 3 are 0.2, 0.2, and 0.32 tCO$_2$/MWh, respectively.

## APPENDIX D

As illustrated in Lemma 1, Fig. D.1 provides corresponding decomposition results of LMCE as energy-dependent and network-dependent components. The decomposition results show that when there is no network congestion, the network-dependent component is 0; however, when the network is congested, the energy-dependent values of LMCE for the entire system are the same, but the network-dependent components of LMCE for different nodes vary. This result indicates that the calculation of LMCE depends on market-clearing outcomes and network operation conditions.

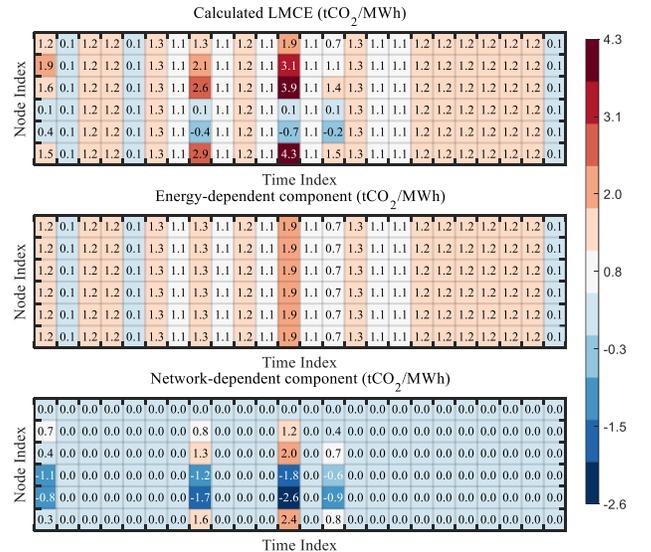

Fig. D.1 The LMCE decomposition result